\documentclass[aps,prb,floatfix,amsmath,amssymb,preprint,nofootinbib]{revtex4-1}
\usepackage{graphicx}
\usepackage{dcolumn}
\usepackage{bm}
\usepackage{epstopdf}
\usepackage{color}

\usepackage{setspace} 
\newcommand{\beq}{\begin{equation}}
\newcommand{\eeq}{\end{equation}}
\newcommand{\beqn}{\begin{eqnarray}}
\newcommand{\eeqn}{\end{eqnarray}}

\usepackage{amsmath}
\usepackage{amssymb}
\begin{document}

\title{Quantum criticality}

\author{Subir Sachdev}
\affiliation{Department of Physics, Harvard University, Cambridge MA
02138}

\author{Bernhard Keimer}
\affiliation{Max-Planck Institute for Solid State Research, Stuttgart, Germany}

\date{\today \\
\vspace{1.6in}}
\begin{abstract}
This is a review of the basic theoretical ideas of quantum criticality, and of their connection
to numerous experiments on correlated electron compounds. A shortened, modified, and edited
version \href{http://dx.doi.org/10.1063/1.3554314}{appeared in the February 2011 issue of Physics Today}. This arxiv version has additional citations
to the literature.
\end{abstract}

\maketitle

\section{Introduction}
\label{sec:intro}

A significant part of modern physics research can be classified as the
study of {\em quantum matter\/}. Its aim is to describe the phases
of large numbers of interacting particles at temperatures low enough so
that quantum mechanics play a crucial role in determining the distinguishing
characteristics. For electrons in solids, the needed ``low''
temperatures can be even higher than room temperature. For gases of trapped atoms, 
ultracold temperatures in the nano-kelvin range are needed.
And in collisions of heavy nuclei at particle accelerator facilities, we obtain a quantum plasma
of quarks  with temperatures approaching those realized soon after the Big Bang.
Remarkably, a common set of ideas on the phases of quantum matter has found application
across this wide range of energy scales. 

This paper will mainly describe the quantum phases of electrons in solids. We will 
develop the theoretical ideas while surveying a variety of recent experiments.

The most common quantum phases of electrons should be familiar to most readers:\\
({\em i\/}) Metals: in the simplest picture, each electron occupies a plane wave state
which extends across the entire solid. Such extended states are responsible for the familiar
electrical conductivity of metals. More precisely, we can characterize each plane wave state
by its momentum, and the quantum state is characterized by the set of momenta of the occupied
states; these momenta have been determined to minimize the total energy. We can then
define a {\em Fermi surface\/} in momentum space, as the surface which separates the regions
of occupied and unoccupied states. The Fermi surface is one of the distinguishing 
characteristics of metal.\\
({\em ii\/}) Superconductors: these are materials which conduct electricity without resistance
below a critical temperature. The electrons bind into pairs known as Cooper pairs. 
Two Cooper pairs, when they are well separated from each other, behave like bosons, and so
at low temperatures the Cooper pairs can `condense' into a state resembling
a Bose-Einstein condensate. The condensate of Cooper pairs is responsible for the ability
to transport charge without dissipation. \\
({\em iii\/}) Insulators: the motion of charge is impeded by the absence of unoccupied
electronic states at nearby energies. Consequently
charge transport is only possible after exciting electrons across
an energy gap, and such processes become very rare at temperatures below the energy gap. 

The above simple classification focused on the motion of the charge carried
by the electron. However, the electron also has a spin, and a study of the
spin configuration in the electron wavefunction allows a more subtle classification of quantum phases. 
A rich variety of magnetic phases are possible, including ferromagnets and antiferromagnets.
These magnetic phases can be either metals, superconductors, or insulators in their
charge sector.

Much of the recent experimental work has focused on ``correlated electron materials''.
Here the electrons can occupy the atomic $d$ or $f$ orbitals, whose smaller spatial
extent increases the importance of the Coulomb repulsion between the electrons.
Although most such materials realize one of the common quantum phases noted above,
they are also amenable to tuning between two or more phases by varying an external
parameter. This parameter could be the pressure
applied on the solid, the value of an external magnetic field, or the density of electrons in the
solid (which can be controlled by the concentration of dopant  ions). Notice that we have
not included the temperature in the list of external parameters: we are studying changes in
the ground state of the electrons, and are not yet interested in the thermal excitations above it. 
We will generically refer to this tuning parameter as $g$.
As $g$ is varied, we have the 
possibility of a {\em quantum phase transition\/} at a critical value $g=g_c$: a qualitative
change in the ground state wavefunction of a large many-body system upon smoothly changing
one or more coupling constants in its Hamiltonian.

Sometimes, the transition between the two phases can be a sudden jump,
which is then a first-order quantum transition, analogous to a first-order
thermally driven phase transition (such as water boiling to steam at atmospheric pressure).
However, more interesting and quite common, is the `continuous' case when the transition
occurs gradually. It is useful to characterize this gradual change by examining the structure of the
wavefunction at different length scales. When we are in a given phase, the wavefunction acquires
the key characteristics of that phase at large length scales. However, there is a length scale, $\xi$, below which
the wavefunction `knows' about the quantum phase transition, and acquires characteristics of the critical point $g=g_c$. 
In a continuous quantum phase transition, the length scale $\xi$ diverges as 
we approach the quantum critical point at $g=g_c$.
The wavefunction for the quantum state at $g=g_c$ is then a complicated superposition
of an exponentially large set of  configurations fluctuating at all length scales: in modern parlance, it has long-range
quantum entanglement.
It is also possible that there
is a critical phase extending over a finite range of $g$  between the 
two traditional phases---we will briefly mention this generalization later.

The quantum critical point $g=g_c$ is defined by the ground state wavefunction,
and so, strictly speaking, it is present only at the absolute zero of temperature ($T$).
Thus, from an experimental perspective, it may seem that a quantum phase transition
is an abstract theoretical idea.
However, as will become amply clear below, the influence of the critical point
extends over a wide regime in the $T>0$ phase diagram: this is the regime
of {\em quantum criticality}, which is crucial for interpreting a wide variety of experiments.

\section{Magnetic insulators}
\label{sec:ins}

We begin our description of quantum phase transitions and quantum criticality
by considering two paradigmatic examples explored in recent experiments.
Both examples are in insulators: so the electron charge is localized, and 
we can focus our attention solely on the orientation
of the electron spins on different sites in the crystal lattice.

\subsection{The quantum Ising chain}

In CoNb$_2$O$_6$ (see Fig~1), only the total electronic spin on the Co$^{2+}$ ion is able to
choose its orientation; all other electrons are paired into spin singlets.
Because of spin-orbit effects, the
Co$^{2+}$ spins have a lower energy 
when their spins are
either parallel or anti-parallel to a given crystalline axis: such spins are referred to
as Ising spins.
\begin{figure}[h]
\includegraphics[width=3in]{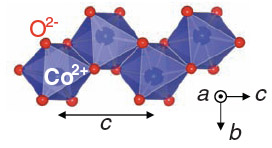}
\caption{A zig-zag chain of Co$^{2+}$ ions, surrounded by oxygen octahedra
in CoNb$_{2}$O$_6$. Coldea {\em et al.} \cite{coldea} showed that 
the spins on Co$^{2+}$ undergo a quantum phase transition
in a magnetic field applied transverse to the chain direction.}
\label{fig:conbo}
\end{figure}
We write the two possible electronic spin states on the Co$^{2+}$ ion at 
site $j$ by $|\uparrow \rangle_j$ and $|\downarrow \rangle_j$.
In the quantum computing terminology, we can say that each Co$^{2+}$ ion
realizes a `qubit'. The spin Hamiltonian of CoNb$_2$O$_6$ has
a coupling between neighboring spins along one-dimensional
zig-zag chains in the crystal so that they prefer to be parallel to each other.
Consequently, in its ground state CoNb$_2$O$_6$ is a ferromagnet, with all spins parallel.
There are two possible ferromagnetic ground states
\begin{equation}
|\Uparrow \rangle = \prod_{j=1}^N |\uparrow \rangle_j \quad \mbox{or} \quad |\Downarrow \rangle = \prod_{j=1}^N |\downarrow \rangle_j \label{ferro}
\end{equation}
where $N$ is total the number of spins on a chain.
The crystal chooses one of these two states depending from small external perturbations. For the crystal Hamiltonian
under study, we note that the choice between these states breaks a spin-inversion symmetry of the Hamiltonian,
under which $|\uparrow \rangle_j$ is interchanged with  $|\downarrow \rangle_j$.

We can now drive a quantum phase transition in CoNb$_2$O$_6$ by applying a magnetic field 
transverse to the preferred crystalline axis, as was done recently
by Coldea {\em et al.}\cite{coldea}. We measure the strength of their magnetic field by $g$.
As $g \rightarrow \infty$, we now see that a ground state very different from (\ref{ferro}) must appear.
As a consequence of the Zeeman coupling, all spins must orient parallel to the applied field;
for a field in $+\hat{x}$ direction, each spin must be an eigenstate of the $x$-component of the spin,
and this leads us to the unique
ground state
\begin{equation}
\left|\Rightarrow \right\rangle = \prod_{j=1}^N |\rightarrow \rangle_j \quad
\mbox{where} \quad |\rightarrow \rangle_j \equiv \frac{|\uparrow \rangle_j + |\downarrow \rangle_j }{\sqrt{2}}. \label{para}
\end{equation}
If we expand out the product in Eq.~(\ref{para}), we obtain an equal superposition of all $2^N$ states of the $N$ qubits: a fact
that is put to good use in quantum computing applications.
Unlike the states in Eq.~(\ref{ferro}), the state in Eq.~(\ref{para}) is invariant under the interchange of
$|\uparrow \rangle_j$ and  $|\downarrow \rangle_j$, and so it 
does not break the spin-inversion
symmetry. It is not a ferromagnet, and we refer to it as a paramagnet. It is important to distinguish
this {\em quantum\/} paramagnet from a thermal paramagnet obtained by simply raising a 
ferromagnet to a high temperature. In the thermal case, the spins are fluctuating in real time in response to incoherent
perturbations from the thermal environment; in contrast, in a quantum paramagnet, 
the spins are in the coherent quantum superposition
of `up' and `down': they are locked into the specific quantum state shown in Eq.~(\ref{para}).

Now, a key point is that it is not possible to vary $g$ and smoothly connect the states in Eq.~(\ref{ferro}),
obtained for $g=0$, to that in Eq.~(\ref{para}), obtained for $g \rightarrow \infty$. To see this, it is crucial to consider the
nature of the $N\rightarrow \infty$ limit. For $N=1$, the two degenerate states in Eq.~(\ref{ferro}) will tunnel into each
other in the presence of any applied transverse field, and so the ground state is, actually, always like Eq.~(\ref{para}).
However, for large $N$, the tunneling amplitude between $|\Uparrow \rangle$ and $|\Downarrow \rangle$ becomes exponentially
small in $N$, because it is the product of the tunneling amplitude of all $N$ spins. Thus, in the thermodynamic limit, the tunneling amplitude is
strictly zero: once a crystal finds itself in (say) the state $|\Uparrow \rangle $, it has an infinite memory of the `up'
orientation, and it will never find the state $|\Downarrow \rangle$. This statement remains true for a finite range of small transverse
fields $g$, when the ground state remains mostly up: the difference between the number of up and down spins measures
the ferromagnetic moment, and the presence of the broken spin inversion symmetry. Thus, there must be a point of
non-analyticity in $g$, where the ferromagnetic moment in the ground state vanishes and 
spin-inversion symmetry is restored: this identifies the quantum critical point at $g=g_c$. 

We can now ask about the evolution of the ground state wavefunction as a function of $g$, between the two limiting
simple cases in Eqs.~(\ref{ferro}) and (\ref{para}). Away from the critical point at $g=g_c$, and at length scales 
larger than $\xi$, the wavefunction effectively `renormalizes' into a structure which is similar to either Eqs.~(\ref{ferro})
or (\ref{para}). In other words, if we describe by the state in terms of `effective' spins representing
the average of the microscopic spins, then the size of the effective spins is $\xi$, and their wavefunction looks like Eq.~(\ref{ferro})
for $g < g_c$, and like Eq.~(\ref{para}) for $g>g_c$. As we noted earlier, $\xi$ diverges as $|g-g_c|$ becomes small, and for length
scales smaller than $\xi$ the wavefunction is like that at $g=g_c$. However, in general, it is not possible to find a simple representation
of this elusive quantum-critical wavefunction. It involves a highly non-trivial quantum superposition of all $2^N$ spin configurations, with correlations between spins which decay with a power-law of distance: consequently, at scales larger than the lattice spacing, it has a scale-invariant structure. The alert reader will notice that $|\Rightarrow \rangle$ also involves a superposition of all $2^N$ states
in the up-down basis: however, it is an equal-weight superposition of all these states, and so can be written as a simple product state
in the left-right basis. In contrast, there is no local basis for which the quantum-critical state at $g=g_c$ takes a simple form: thus
it is highly quantum `entangled'.

\subsection{Dimer antiferromagnet}

\begin{figure}
\centering
 \includegraphics[width=4.5in]{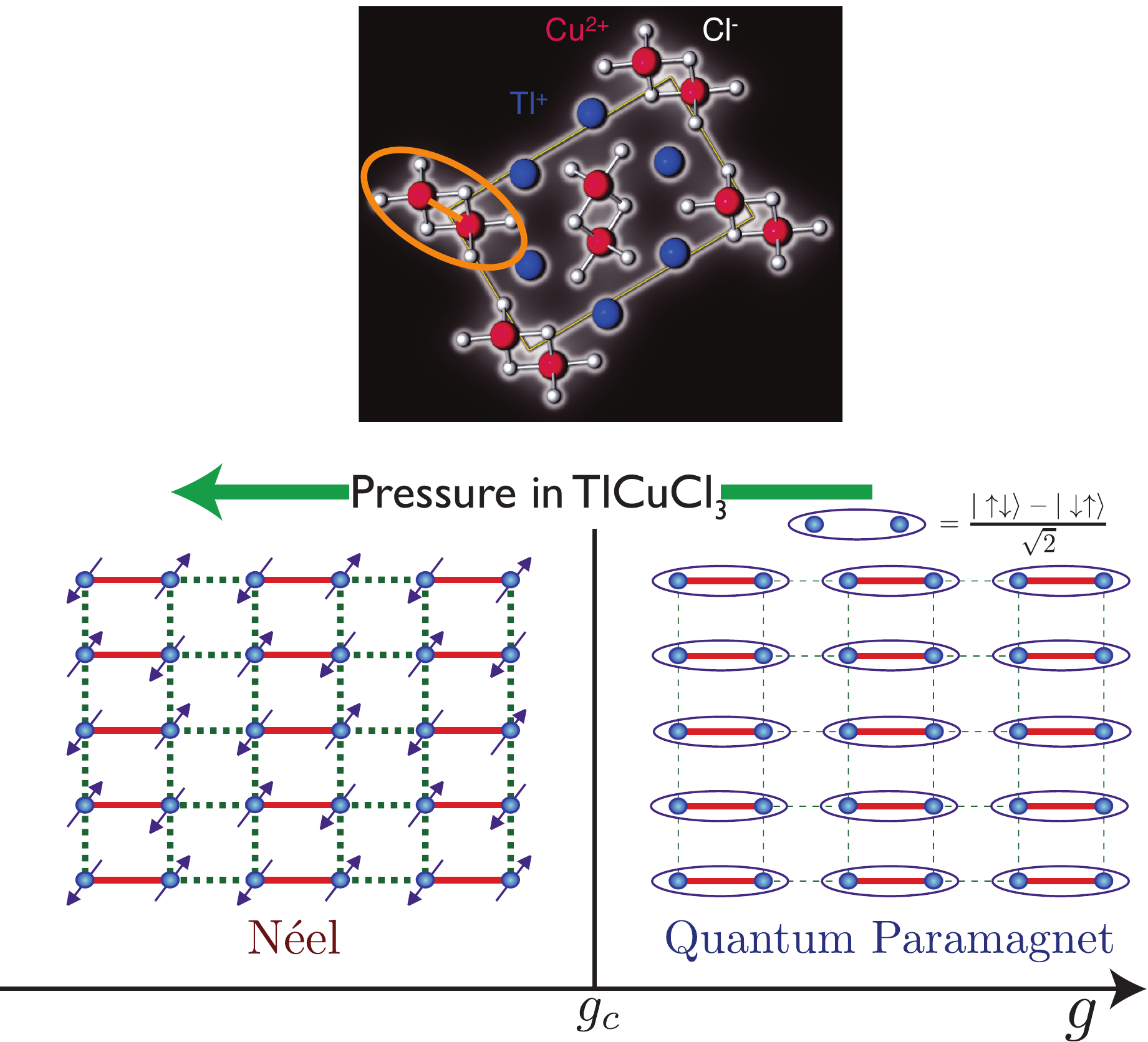}
 \caption{Top: The unit cell of the insulating antiferromagnet TlCuCl$_3$. Bottom: A simple realization of the dimer 
 antiferromagnet on the square lattice.
The full red lines represent an exchange interaction $J>0$, while the dashed green lines have exchange $J/g$ with $g>1$. The ellipses represent a singlet valence
 bond of spins $(|\uparrow \downarrow \rangle - | \downarrow \uparrow \rangle )/\sqrt{2}$.}
\label{fig:dimer}
\end{figure}

We turn to our second example: the dimer antiferromagnet, which is realized in TlCuCl$_3$, and which will have a closer
connection to the materials considered later in this paper.
This material has unpaired, localized electron spins on each Cu$^{2+}$ ion. We introduce the spin $S=1/2$ operator ${\bf S}_j$ which
acts on site $j$, and  exchange couplings, $J_{ij}$, between the Cu ions; the spins are then 
described by the Hamiltonian
\begin{equation}
H = \sum_{i<j} J_{ij} {\bf S}_i  \cdot  {\bf S}_j .
\label{eq:HJ}
\end{equation}
A key feature is that $J_{ij} > 0$ {\em i.e.} the exchanges are antiferromagnetic, and prefer {\em anti\/}-parallel spins (unlike the inter-spin couplings of the Ising chain above).
Further, they are dimerized: each spin $j$ is coupled to only a single partner spin with the largest exchange $J$, and all other
exchanges are smaller. A simple example of a dimerized antiferromagnet is illustrated in Fig.~\ref{fig:dimer}, which has exchanges
$J$ and $J/g$ with $g > 1$.

We describe the quantum phase transition in the model of Fig.~\ref{fig:dimer}. At $g=1$, the ground state has N\'eel (or antiferromagnetic) order with the spins polarized in a checkerboard pattern as shown in the figure: this state breaks the spin rotation symmetry of the Hamiltonian.
This is the analog of the ferromagnetic state of the Ising chain, except that now the spins are polarized in a staggered spatial
pattern. With the staggering, each up spin has down neighbors (and vice versa), and so each term in the exchange energy
of $H$ is negative.

Now, let us contrast this with the very different ground state for $g \rightarrow \infty$. Notice that at $g=\infty$, the couplings
between different dimers vanish, and the Hamiltonian decouples into a sum over independent pairs of spins on each dimer.
We can easily find the ground state for each dimer of spins: it is the singlet valence bond: $(|\uparrow \downarrow \rangle - | \downarrow \uparrow \rangle )/\sqrt{2}$. A key property of this singlet bond is that it is rotationally invariant; although the two spins
are always anti-parallel, they are equally likely to point in any direction in spin space. The ground state of $H$ at $g=\infty$
is then a product over such singlet valence bonds, as illustrated in Fig.~\ref{fig:dimer}. 

The remaining discussion for $H$ then parallels that for the Ising chain. There is N\'eel order for a range of $g$ with broken spin rotation symmetry. Spin rotation symmetry is restored at a quantum critical point $g=g_c$, beyond which the ground state 
is a spin-singlet quantum paramagnet. There is a continuous transition at $g=g_c$, where the ground state has non-trivial, 
entanglement between the spins at all length scales.

\begin{figure}
\centering
 \includegraphics[width=4in]{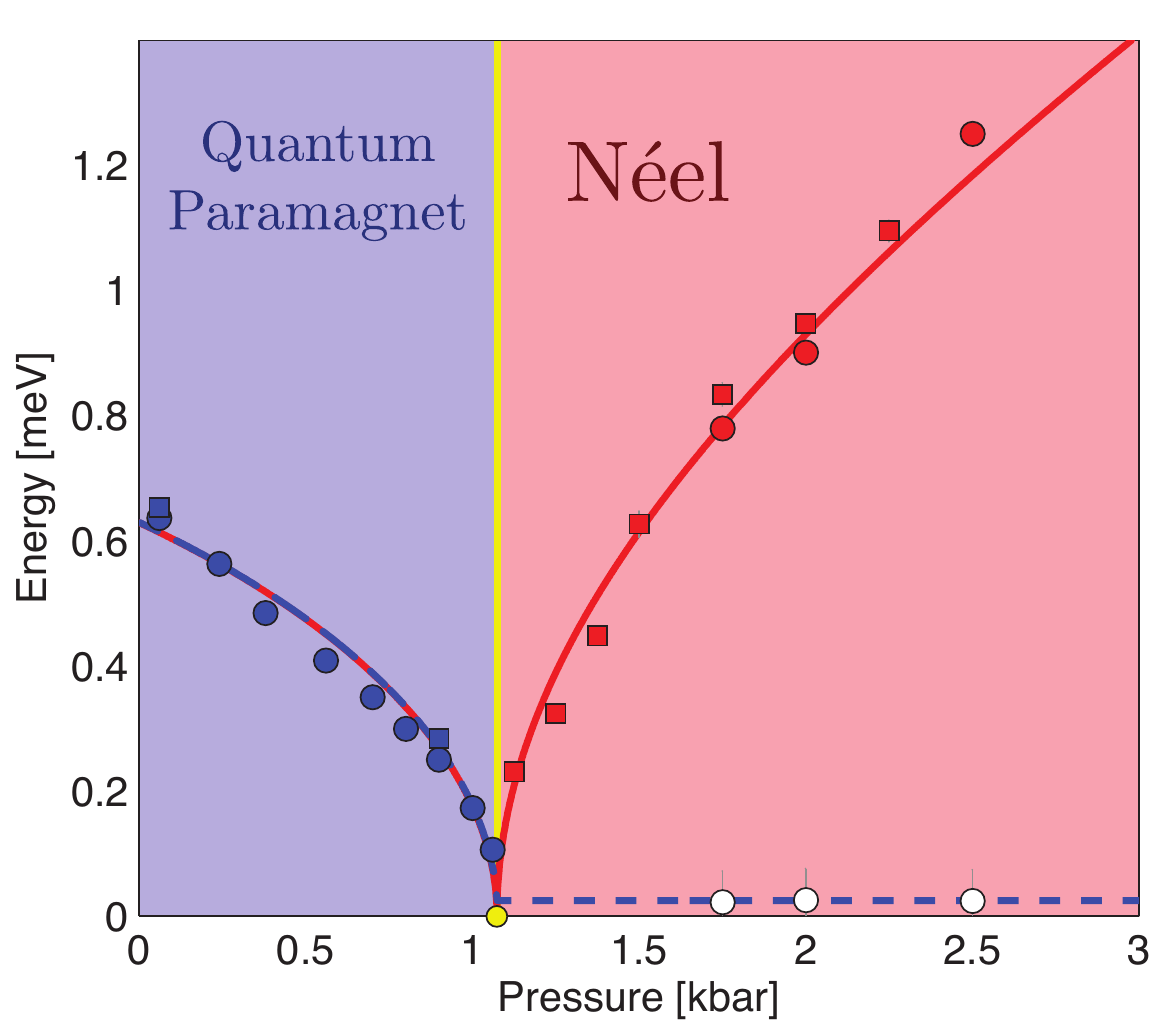}
 \caption{Observation of the quantum phase transition in TlCuCl$_3$ in neutron scattering experiments by Ruegg {\em et al.} \cite{ruegg}.
 At low pressure, the ground state is the $g>g_c$ quantum paramagnet. Its excitations are `triplons' created by breaking
nearest-neighbor singlet bonds (see Fig.~\ref{fig:qc}), and these are represented by the blue symbols; their energy vanishes as
the critical point (QCP) is approached. At high pressure we have the N\'eel state, which has antiferromagnetic order
at low temperatures. Here, there are two types of excitations: the `spin waves' are slow deformations
of the N\'eel order (see Fig.~\ref{fig:qc}) which have vanishing energy in the limit of long wavelengths, and these are represented by the white circles. The other excitation is an analog of the Higgs boson: it is an oscillation in the magnitude of the local N\'eel order, and its energy
also vanishes upon approaching the critical point (red symbols).}
\label{fig:ruegg}
\end{figure}

The distinct ground states of Fig.~\ref{fig:dimer}, and their very different spectra of excitations have been
detected by neutron scattering experiments on TlCuCl$_3$. The value of $g$ is tuned by applied pressure,
and the results are shown in Fig.~\ref{fig:ruegg}.

\section{Quantum criticality}

We turn, finally, to the experimental significance of the isolated quantum critical state at $g=g_c$
upon which we have lavished much attention above. For this, we need to consider the influence
of a non-zero temperature on the quantum phase transition in the ground state. As we will now see,
the transition leaves a clear fingerprint on a large portion of the $T>0$ phase diagram \cite{chn}.

\begin{figure}
\centering
 \includegraphics[width=4.5in]{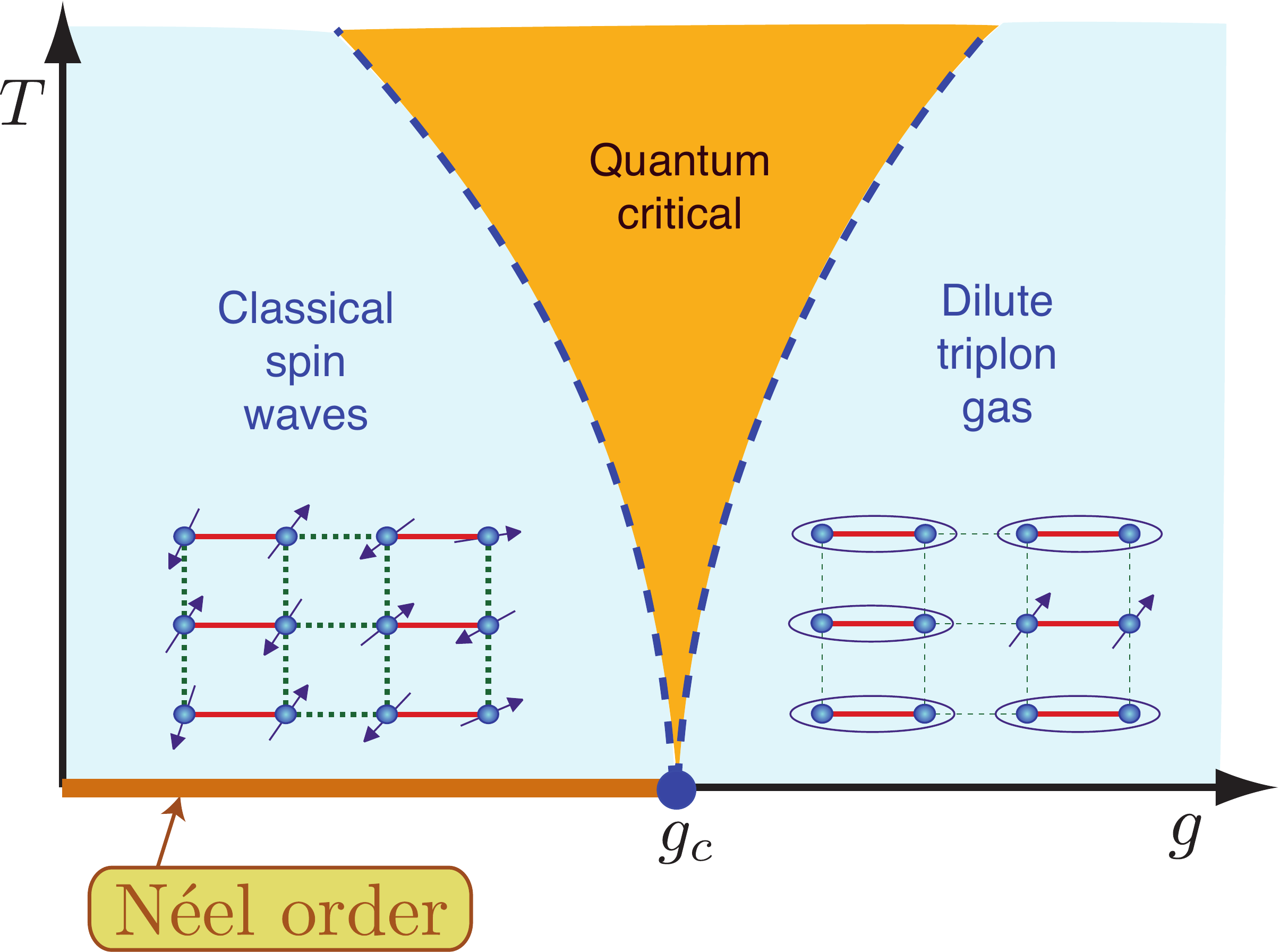}
 \caption{Non-zero temperature ($T$) phase diagram for the model in Fig.~\ref{fig:dimer}.
 For $g<g_c$, the low energy excitations are slow deformations of the N\'eel order, the spin waves.
 These have strong non-linear couplings which destroy the N\'eel order at all $T>0$, and the long time dynamics
 have a classical description in a theory of interacting spin waves. For $g>g_c$, the excitations are the triplet $S=1$ excited states
 of each dimer. These become mobile and form a dilute gas of `triplons' whose dynamics can be described by the classical
 Boltzmann equation. Quantum criticality appears in the intermediate orange region, where there is no effective classical theory at the
 scale of the characteristic spin equilibration time; instead we have the strongly coupled dynamics of the non-trivial critical excitations
 which have neither a particle or wave interpretation.}
\label{fig:qc}
\end{figure}

We begin by sketching the phase diagram in the $T$ and $g$ plane for the dimer antiferromagnet
of Fig.~\ref{fig:dimer}.
This is shown in Fig.~\ref{fig:qc}.

At $T=0$, we have the evolution in the ground state as described in the previous section. 
For a small non-zero $T$, and on either side of $g_c$, we expect that a reasonable description can be achieved
by mixing in a few low-lying excited states above the ground state. This is indeed the case in the 
 blue-colored regions of Fig.~\ref{fig:qc}. For $g<g_c$, the excitations are spin-waves: slow wave-like
 deformations of the N\'eel-ordered state found in the ground state. On the other hand, for $g>g_c$ the excitations
 are the ``triplon'' particles: as illustrated in Fig.~\ref{fig:qc}, these are local triplet excitations of the singlet bonds
 on the dimer, which hop from dimer to dimer. Moreover, a key observation is that the dynamics of these waves or particles
can be described by an effective classical theory.

Let us make the last statement a bit more precise. Imagine perturbing the system away from thermal
equilibrium, and then observing that it takes a time $\tau_{\rm eq}$ to recover to a state with local equilibrium.
Then, the dynamics at the time scales of order $\tau_{\rm eq}$ can be accurately described by models
of interacting classical waves or particles.

Now we turn to orange quantum critical region of Fig.~\ref{fig:qc}. Here, at short length and time scales, the system
acts as if it is in the quantum critical ground state at $g=g_c$. This is similar to our earlier observation that the ground
state wavefunction looks like that one of the phases adjacent to the critical point only at length scales larger than $\xi$.
Upon moving to longer scales, it discovers thermal effects
{\em before} it realizes that the coupling $g$ may not be exactly $g_c$ {\em i.e.\/} before it reaches $\xi$. 
As thermal effects are stronger at larger $T$, quantum criticality is realized over a
fan which {\em widens\/} (somewhat counter-intuitively) 
with increasing $T$. Because the system never learns about the length scale $\xi$ here,
we may as well as set $g=g_c$ in the entire quantum critical region.
We have neither the simple particle-like or wave-like excitations
of the two non-critical states of Fig.~\ref{fig:dimer}, but the strongly-coupled dynamics of the quantum critical point
and its excitations which entangle the microscopic quantum degrees of freedom in a non-trivial manner. 

A key characteristic of quantum criticality is its thermal equilibration time \cite{sachdev}
\begin{equation}
\tau_{\rm eq} = \mathcal{C}_{\rm eq} \frac{\hbar}{k_B T},  \label{qc}
\end{equation}
where $\hbar$ is Planck's constant, $k_B$ is Boltzmann's constant, and $\mathcal{C}_{\rm eq}$ is a
dimensionless universal number {\em i.e.\/} remarkably, $\mathcal{C}_{\rm eq}$ is independent of the specific
microscopic form of the Hamiltonian (there are exceptions to Eq.~(\ref{qc}) for quantum critical
points which don't have strong interactions). Also the value of $\tau_{\rm eq}$ in Eq.~(\ref{qc})
is the shortest possible: the blue regions of Fig.~\ref{fig:qc}
have $ \tau_{\rm eq} \gg \hbar/(k_B T)$, which is the inequality needed for an
effective classical description. No effective classical model provides an accurate description in
the orange region of quantum
criticality with Eq.~(\ref{qc}), and quantum effects have to be retained for an understanding
of the entangled excitations of the critical state and their complex interactions. 
Given its smallest value of $\tau_{\rm eq}$,
quantum criticality realizes the {\em perfect fluid\/} \cite{kss}.

\begin{figure}
\centering
 \includegraphics[width=4in]{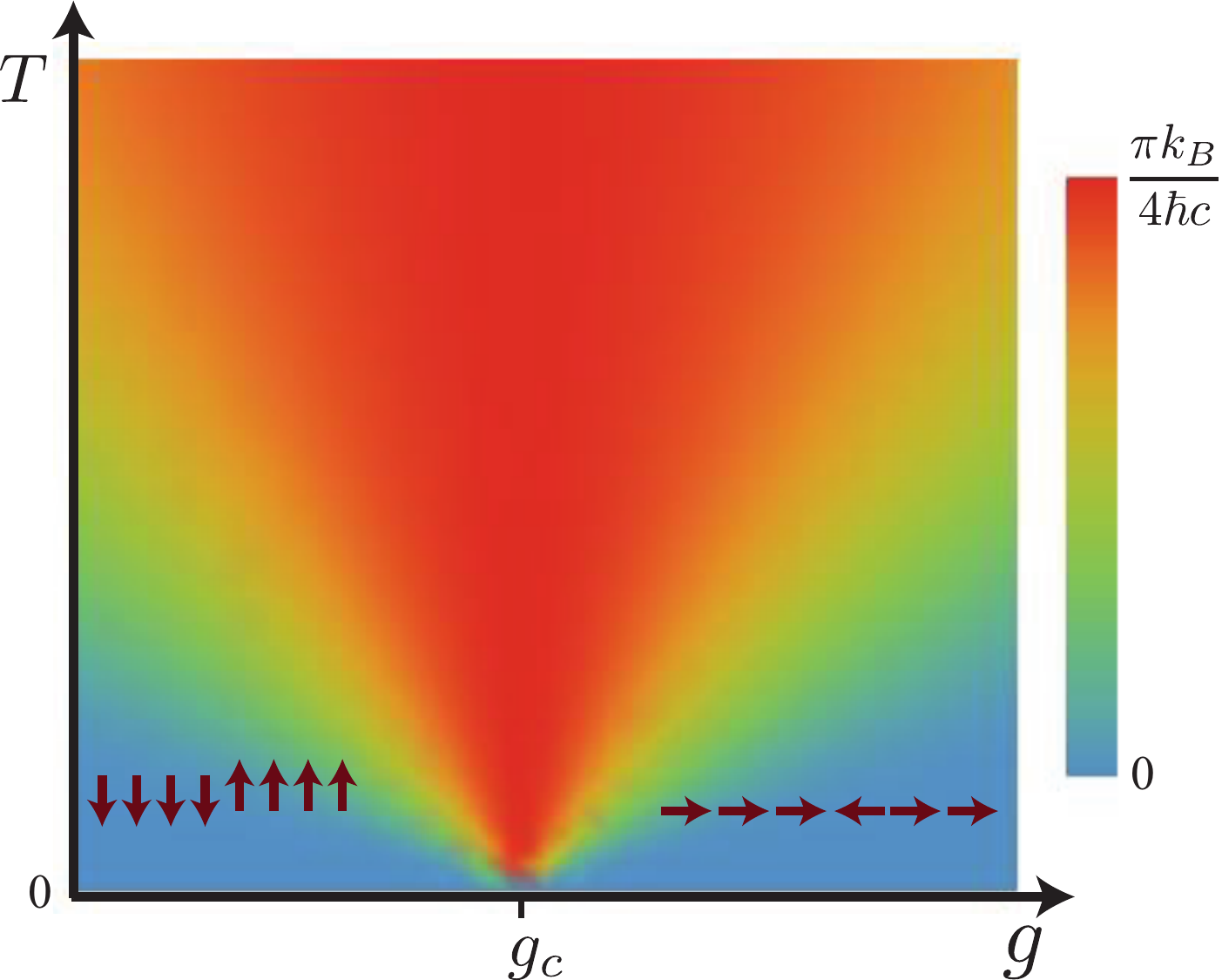}
 \caption{Non-zero temperature properties of the Ising quantum spin chain which models CoNb$_2$O$_6$ shown in
 Fig.~\ref{fig:conbo}. Shown are theoretical computations from the exactly solvable spin chain with nearest-neighbor exchange.
 The color plot indicates the value of the $(4\hbar c /\pi k_B) (d \xi^{-1} /dT)$, where $\xi$ is the spin correlation length and $c$ is the velocity
 of spin excitations; this dimensionless number has a $T$ dependence similar to that of the $T$ derivative of $\tau_{\rm eq}^{-1}$
 of non-integrable strongly-interacting quantum critical points. 
 Also indicated are typical spin configurations in the two low temperature regimes. For $g<g_c$, we have the ferromagnetic configurations
 of Eq.~(\ref{ferro}) separated by domain walls, while for $g>g_c$ we have the paramagnetic state of Eq.~(\ref{para}) with
 its `reversed spin' excitations; here $|\rightarrow \rangle = (|\uparrow \rangle + |\downarrow \rangle)/\sqrt{2}$
 and $|\leftarrow \rangle = (|\uparrow \rangle - |\downarrow \rangle)/\sqrt{2}$.}
\label{fig:ising}
\end{figure}

We also illustrate the similar $T>0$ crossovers for the Ising chain found in CoNb$_2$O$_6$ in Fig.~\ref{fig:ising}.
The quantum spin chain with only nearest-neighbor exchange couplings is exactly solvable, and we plot
a quantity closely related to the temperature derivative of $\tau_{\rm eq}^{-1}$ for generic quantum critical points: 
these clearly illustrate the 3 regimes of
Fig.~\ref{fig:qc}, including the central regime of quantum criticality.

The behavior in Eq.~(\ref{qc}) can be detected in experiments by measuring various
response functions as a function of both frequency ($\omega$) and $T$.
Then we expect \cite{varma} these results to depend only upon the single variable $\hbar \omega /k_B T$.
Inelastic neutron scattering experiments on insulating compounds with spin-1/2 ions on one-dimensional and geometrically frustrated two-dimensional lattices \cite{helton}, as well as metallic copper oxides \cite{keimer} and heavy-fermion compounds \cite{schroder} near the doping levels at which antiferromagnetic long-range order vanishes have revealed that this function scales with the ratio $\omega /T$.

The transport properties of the quantum critical region also enjoy a great deal of universality.
This is expected from our reasoning above, because the values of the transport coefficients
depend on the same processes which establish local equilibrium.
We mention here the example that appeared in a quantum-critical description of momentum transport in
the quark-gluon plasma \cite{kss}, but applies more generally (in lattice models, we require that momentum
degradation by `umklapp' scattering is subdominant). We can characterize momentum
transport by a shear viscosity $\eta$, which then obeys
\begin{equation}
\frac{\eta}{s} = \mathcal{C}_{m} \frac{\hbar}{k_B}
\end{equation}
where $s$ is the entropy density and $\mathcal{C}_{m}$ is another universal dimensionless number of order unity.
It has so far not been possible to measure the value of $\eta$ for interacting electrons in quantum matter.
However, such measurements are more natural in ultracold atoms and the quark-gluon plasma \cite{shuryak} (see articles in the
May 2010 issue of {\em Physics Today}.). The universality of quantum critical transport extends also 
to charge and spin transport, and their associated conductivities.

In some simpler cases, mainly in insulators, the quantum critical point $g=g_c$ is described by a
`conformal field theory' (CFT). These are quantum field theories well-known in statistical mechanics
and string theory, which enjoy a large group of spacetime symmetries, including relativistic
and scale invariance. In particular, the simple square lattice model in Fig.~\ref{fig:dimer}
with the Hamiltonian in Eq.~(\ref{eq:HJ}) has a critical point described by a CFT, in which the
velocity of `light' is replaced by the spin wave velocity. Our analytic understanding of quantum criticality
is most advanced for those critical points described by CFTs. This is partly due to the
AdS/CFT correspondence discovered by string theorists, which relates quantum criticality to dynamics near the horizon
of a black hole in an anti-de Sitter (AdS) space with one more spatial dimension than the CFT.
Remarkably, the characteristic quantum critical time in Eq.~(\ref{qc}) is mapped onto the damping
time of quasi-normal modes of various fields, including gravity, around the black hole, with $T$ mapped onto 
the Hawking temperature of the black hole. In the past two years, there has also been much research activity
extending such string theory-inspired ideas to quantum criticality not associated with a
relativistic CFT, and some very promising results have been obtained \cite{mcgreevy}.

\section{Quantum phase transitions of metals}

\begin{figure}
\centering
 \includegraphics[width=4.5in]{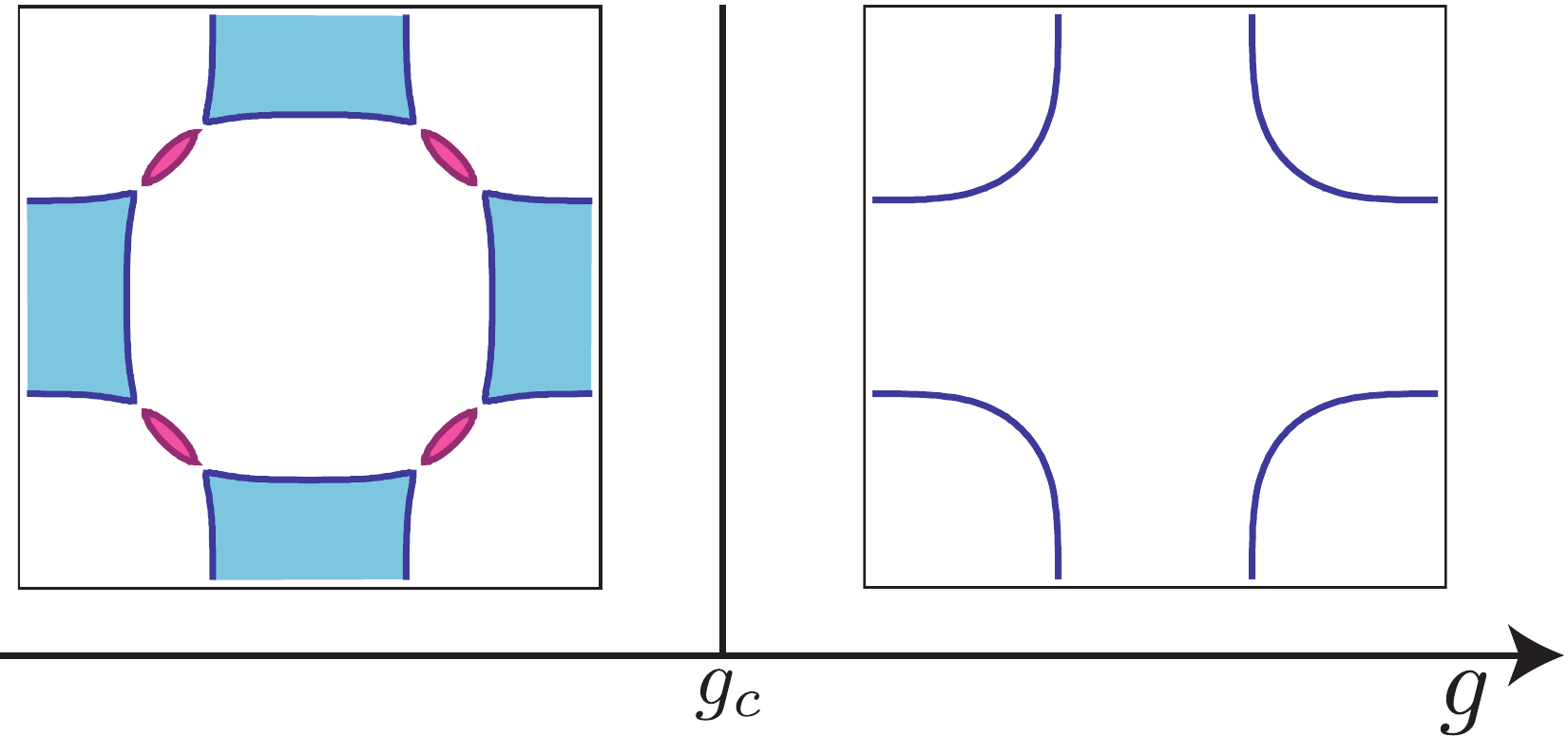}
 \caption{Evolution of the Fermi surface across a quantum phase transition in a metal involving the
 onset of antiferromagnetism. The spin ordering is similar to that in Fig.~\ref{fig:dimer}, but takes place in a metal
 rather than in an insulator. We have chosen a Brillouin zone and Fermi surface appropriate for the
 electron-doped cuprate superconductors, where the value of the coupling $g$ is tuned
 by changing the electron concentration. The paramagnetic phase for $g>g_c$ has a `large' Fermi surface,
 which is determined mainly by the underlying band structure. For $g<g_c$, the antiferromagnetic order
 appears, and the diffraction of the electron waves off this order modifies their dispersion, and leads
 to the small electron (blue) and hole (red) `pocket' Fermi surfaces. The quantum critical point now
 involves a non-trivial interplay between quantum fluctuations associated with the Fermi surface evolution
 and the antiferromagnetic order.}
\label{fig:sdw}
\end{figure}

Now let us consider the same transition as in Fig.~\ref{fig:dimer}, between an antiferromagnet and a paramagnet,
but in a metal. As noted earlier, the distinguishing characteristic of a metal is its Fermi surface, the demarcation
between the occupied and unoccupied electronic states in momentum space. 
The appearance of N\'eel order for $g<g_c$ is expected
to change the electronic dispersion spectrum, and so we expect that the quantum phase transition should
be accompanied by a qualitative change in the Fermi surface. This is often the case, and is illustrated in
Fig.~\ref{fig:sdw}.

The change in Fermi surface has a profound effect on the nature of the quantum critical point at $g=g_c$.
In the insulator, this was described by a CFT which focused primarily on the fluctuations of the N\'eel
order at long wavelengths. Now the CFT no longer applies, as we have to include the fermionic
excitations in the vicinity of the Fermi surface \cite{hmm}, which is itself undergoing a topological change.
The resulting theory has been shown to be strongly coupled in two spatial dimensions \cite{max}, and
unravelling its critical properties is a key challenge for theoretical studies of quantum criticality.

\begin{figure}
\includegraphics[width=6.5in]{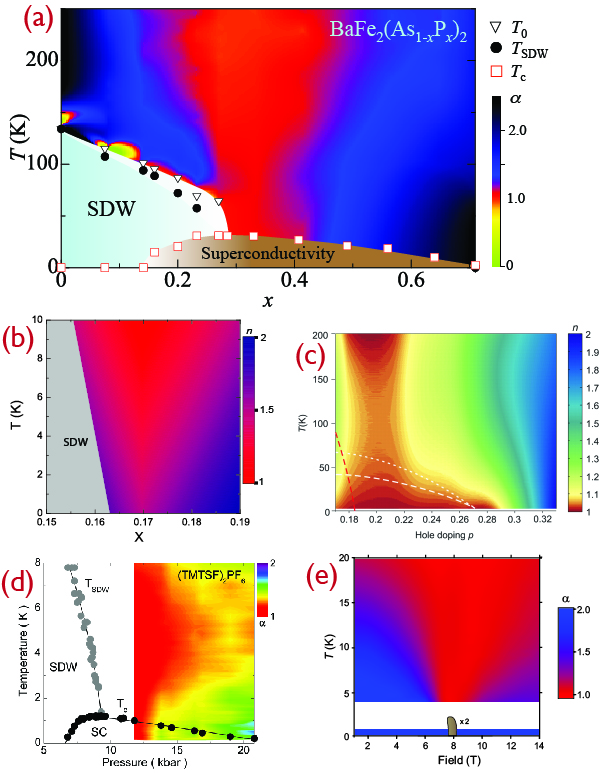}
\caption{Temperature exponent of the resistivity of the pnictide compound \cite{kasahara} (top) BaFe$_2$(As$_{1-x}$P$_x$)$_2$, the ruthenate \cite{grigera} (middle left) Sr$_3$Ru$_2$O$_7$, 
the cuprate \cite{cooper}
(middle right) La$_{2-x}$Sr$_{x}$CuO$_4$ in a magnetic field high enough to suppress superconductivity,
and the organic \cite{doiron} (TMTSF)$_2$PF$_6$.
The color plots show the local value of the resistivity exponent $\alpha$ (or $n$) equal to $d \ln \rho/d \ln T$.
For the pnictide, the region labeled SDW (spin density wave) region has antiferromagnetic order.
For Sr$_3$Ru$_2$O$_7$, the small finger, appearing at low $T$ near the critical field, is a region where
the point group symmetry is broken and Ising-nematic order appears (see Section~\ref{sec:compete}).}
\label{fig:transport}
\end{figure}

In experimental studies of correlated-electron materials, much attention has
focused on the so-called {\em strange metal\/} behavior. One of the hallmarks of the strange
metal is that the electrical resistivity, $\rho$, is proportional to $T$. The behavior of this elementary transport
quantity is in stark contrast to the predictions of the standard Fermi-liquid theory of metals,
according to which $\rho \sim T^2$ at low temperatures. A small sample from the many examples
of strange-metal behavior is shown in Fig.~\ref{fig:transport}.

Given the strong similarity between the structures of the crossovers in Fig.~\ref{fig:transport} and~\ref{fig:qc},
it is natural to associate the strange metal with quantum criticality.
Indeed, by taking advantage of external control parameters, a diverse variety of metallic materials including copper oxides \cite{cooper}, ruthenium oxides \cite{grigera}, iron pnictides \cite{kasahara}, organic metals \cite{doiron}, and
heavy-fermion compounds \cite{gegenwart} has been tuned to quantum critical points. Remarkably, in all of these cases the electrical resistivity, $\rho$, was found to be strictly proportional to temperature precisely at the quantum critical point, whereas subtle deviations from $T$-linear behavior were observed away from it. As an example, consider the recently discovered pnictide high temperature superconductor BaFe$_2$(As$_{1-x}$P$_x$)$_2$. Here the
strange metal coincides precisely with the onset of antiferromagnetism. Thus it is natural to associate the strange metal
with the quantum criticality of the Fermi surface transformation like that in Fig.~\ref{fig:sdw}.

In the materials with the highest-temperature superconductivity, the hole-doped cuprates, the identification of the strange metal
with a quantum-critical Fermi surface change (similar to that in Fig.~\ref{fig:sdw})
was made most explicitly by Taillefer and collaborators \cite{doiron}. For the cuprates with the highest superconducting critical temperatures,
such a proposal does not seem tenable at first sight,
because these materials have no evident magnetic ordering transition at the dopings
where strange metal behavior is dominant. However, a strong magnetic field has been applied to suppress
the superconductivity and reveal the strange metal at low temperatures, and this magnetic field could also shift the position of the magnetic
ordering critical point \cite{moon}.
Strong supporting evidence for this picture is provided by the remarkable observation of Fermi pockets by quantum oscillation experiments carried out in ultrapure
crystals in magnetic fields reaching record laboratory strength \cite{louis}: these oscillations suggest a Fermi surface structure
similar to that in the left panel of Fig.~\ref{fig:sdw}, and so are consistent with the quantum critical point $g=g_c$
being where needed for the strange metal behavior.

We note that in some of the hole-doped cuprate superconductors, the strange-metal region appears to extend over a wider regime of parameters
at low $T$ and high magnetic fields (see Fig.~\ref{fig:transport}). Similar observations have been made in organic metals \cite{doiron}.
This could indicate the existence of an intermediate quantum critical {\em phase\/} at high fields,
or be associated with a shift of the antiferromagnetic quantum critical point due to the onset of 
superconductivity \cite{moon}.

\section{Competing orders}
\label{sec:compete}

Let us now return to the layout of the phase diagrams of Fig.~\ref{fig:transport}.
Notice that all of them include additional electronic phases, in contrast to the `bare' quantum critical point in the phase diagram of the simple insulator shown in Fig.~\ref{fig:ruegg}. The most prominent example is the Cooper pairing instability leading to unconventional  superconductivity in the cuprates and the pnictides. By `unconventional' we mean that the wavefunction of the electrons
in a Cooper pair has a non-trivial spatial dependence, and changes sign between different regions
of the Brillouin zone in momentum space.
Both the shape of the phase boundaries and the symmetry of the superconducting order parameters suggest that the pairing instabilities in these materials are driven by fluctuations of the order parameter that characterizes the quantum critical point, namely antiferromagnetic spin fluctuations \cite{scalapino,fink,mazin,kuroki,jiangping,fermiology}. 
One of the virtues of the theory of superconductivity mediated by antiferromagnetic spin fluctuations is that it yields
an immediate and simple explanation for the unconventional superconductivity: the electrons bind into Cooper pairs only if the
the pair wavefunction has opposite signs on patches of the Fermi surface which are connected by the antiferromagnetic ordering
wavevector. 
A major open question in the field is whether such a conventional description of a static Fermi surface coupled to a pairing boson
(the antiferromagnetic spin fluctuations) can yield the complete story of the pairing mechanism. 
At least in the more strongly interacting systems, like the cuprates,
it appears that non-Fermi liquid effects, associated perhaps with
critical fluctuations of the Fermi surface geometry, are essential for a complete description.

One of the difficulties in answering this central question is the presence of additional instabilities in the 
charge channel \cite{fermiology,YK00,HM00,OKF01}, 
including in particular the `electronic nematic' state. In this state, the point-group symmetry of the Hamiltonian describing the electron system is spontaneously broken in a manner analogous to liquid-crystalline order in complex fluids. This state was first identified \cite{borzi} in Sr$_3$Ru$_2$O$_7$ (Fig.~\ref{fig:transport}), and there is evidence for substantial nematic correlations also in the cuprates  and the pnictides 
close to their antiferromagnetically ordered phases \cite{hinkov}. In the presence of quenched disorder, these and other ordering phenomena can coexist in an inhomogeneous manner, which may explain some of the conflicting reports in the extensive experimental literature on these materials. In any case, it is clear, from theory and experiment, that both pairing and such charge channel instabilities are ubiquitous
features of the vicinity of the magnetic quantum criticality of Fermi surfaces in two spatial dimensions. 

For the future, we need a comprehensive theory of these multiple-ordering phenomena, and their interplay with fermionic excitations
near the Fermi surface. In particular, a full understanding of ``non-linear'' effects between various orders
is needed: how one static or fluctuating order, enhances or suppresses other orders. This will help
describe the phase diagrams in a strong magnetic field, which have emerged as a key test of theoretical
proposals \cite{sswhere}.
The rapid progress that has been made over the past few years indicates
that we are well on our way towards a systematic description of quantum criticality and competing orders in metals with strong electronic correlations.

\end{document}